


\font\titlefont = cmr10 scaled\magstep 4
\font\sectionfont = cmr10
\font\littlefont = cmr5
\font\eightrm = cmr8

\def\ss{\scriptstyle}
\def\sss{\scriptscriptstyle}

\newcount\tcflag
\tcflag = 0  
\ifnum\tcflag = 0 \magnification = 1200 \fi  

\global\baselineskip = 1.2\baselineskip
\global\parskip = 4pt plus 0.3pt
\global\abovedisplayskip = 18pt plus3pt minus9pt
\global\belowdisplayskip = 18pt plus3pt minus9pt
\global\abovedisplayshortskip = 6pt plus3pt
\global\belowdisplayshortskip = 6pt plus3pt


\def\endignore{}
\def\ignore #1\endignore{}

\newcount\dflag
\dflag = 0


\def\monthname{\ifcase\month
\or January \or February \or March \or April \or May \or June%
\or July \or August \or September \or October \or November \or December
name \fi}

\newcount\dummy
\newcount\minute  
\newcount\hour
\newcount\localtime
\newcount\localday
\localtime = \time
\localday = \day

\def\advanceclock#1#2{ 
\dummy = #1
\multiply\dummy by 60
\advance\dummy by #2
\advance\localtime by \dummy
\ifnum\localtime > 1440 
\advance\localtime by -1440
\advance\localday by 1
\fi}

\def\settime{{\dummy = \localtime%
\divide\dummy by 60%
\hour = \dummy
\minute = \localtime%
\multiply\dummy by 60%
\advance\minute by -\dummy
\ifnum\minute < 10
\xdef\spacer{0} 
\else \xdef\spacer{}
\fi %
\ifnum\hour < 12
\xdef\ampm{a.m.} 
\else
\xdef\ampm{p.m.} 
\advance\hour by -12 %
\fi %
\ifnum\hour = 0 \hour = 12 \fi
\xdef\timestring{\number\hour : \spacer \number\minute%
\thinspace \ampm}}}



\def\endtitle{}
\def\title#1\endtitle{\vskip.5in\titlefont
\global\baselineskip = 2\baselineskip
#1\vskip.4in
\baselineskip = 0.5\baselineskip\rm}

\def\endauthors{}
\def\authors#1\endauthors{#1}

\def\endabstract{}
\def\abstract#1\endabstract{\vskip .3in%
\centerline{\sectionfont\bf Abstract}%
\vskip .1in
\noindent#1}

\newcount\nsection
\newcount\nsubsection

\def\section#1{\global\advance\nsection by 1
\nsubsection=0
\bigskip\noindent\centerline{\sectionfont \bf \number\nsection.\ #1}
\bigskip\rm\nobreak}

\def\subsection#1{\global\advance\nsubsection by 1
\bigskip\noindent\sectionfont \sl \number\nsection.\number\nsubsection)\
#1\bigskip\rm\nobreak}


\def\appendix#1#2{\bigskip\noindent%
\centerline{\sectionfont \bf Appendix #1.\ #2}
\bigskip\rm\nobreak}


\newcount\nref
\global\nref = 1

\def\ref#1#2{\xdef #1{[\number\nref]}
\ifnum\nref = 1\global\xdef\therefs{\noindent[\number\nref] #2\ }
\else
\global\xdef\oldrefs{\therefs}
\global\xdef\therefs{\oldrefs\vskip.1in\noindent[\number\nref] #2\ }%
\fi%
\global\advance\nref by 1
}

\def\listrefs{\vfill\eject\section{References}\therefs}


\newcount\nfoot
\global\nfoot = 1

\def\foot#1#2{\xdef #1{(\number\nfoot)}
\footnote{${}^{\number\nfoot}$}{\eightrm #2}
\global\advance\nfoot by 1
}


\newcount\nfig
\global\nfig = 1

\def\fig#1{\xdef #1{(\number\nfig)}
\global\advance\nfig by 1
}


\newcount\cflag
\newcount\nequation
\global\nequation = 1
\def\eqlabel{(1)}

\def\nexteqno{\ifnum\cflag = 0
\global\advance\nequation by 1
\fi
\global\cflag = 0
\xdef\eqlabel{(\number\nequation)}}

\def\lasteqno{\global\advance\nequation by -1
\xdef\eqlabel{(\number\nequation)}}

\def\label#1{\xdef #1{(\number\nequation)}
\ifnum\dflag = 1
{\escapechar = -1
\xdef\draftname{\littlefont\string#1}}
\fi}

\def\clabel#1#2{\xdef\eqlabel{(\number\nequation #2)}
\global\cflag = 1
\xdef #1{\eqlabel}
\ifnum\dflag = 1
{\escapechar = -1
\xdef\draftname{\string#1}}
\fi}

\def\cclabel#1#2{\xdef\eqlabel{#2)}
\global\cflag = 1
\xdef #1{\eqlabel}
\ifnum\dflag = 1
{\escapechar = -1
\xdef\draftname{\string#1}}
\fi}


\def\eeq{}

\def\eqnn #1\eeq{$$ #1 $$}

\def\eq #1\eeq{
\ifnum\dflag = 0
{\xdef\draftname{\ }}
\fi 
$$ #1
\eqno{\eqlabel \rlap{\ \draftname}} $$
\nexteqno}



\def\eol{& \eqlabel \rlap{\ \draftname} \crcr
\nexteqno
\xdef\draftname{\ }}

\def\eeol{& \eqlabel \rlap{\ \draftname}
\nexteqno
\xdef\draftname{\ }}

\def\eolnn{\cr
\global\cflag = 0
\xdef\draftname{\ }}

\def\eeolnn{\xdef\draftname{\ }}

\def\eqa #1\eeq{
\ifnum\dflag = 0
{\xdef\draftname{\ }}
\fi 
$$ \eqalignno{ #1 } $$
\global\cflag = 0}


\def\ie{{\it i.e.\/}}

\def\apriori{{\it a priori\/}}


\def\ijmp#1#2#3{{\it Int.~J.~Mod.~Phys.} {\bf A#1} (19#2) #3}

\def\npb#1#2#3{{\it Nucl.~Phys.} {\bf B#1} (19#2) #3}
\def\plb#1#2#3{{\it Phys.~Lett.} {\bf #1B} (19#2) #3}

\def\prd#1#2#3{{\it Phys.~Rev.} {\bf D#1} (19#2) #3}
\def\pr#1#2#3{{\it Phys.~Rev.} {\bf #1} (19#2) #3}

\def\prl#1#2#3{{\it Phys.~Rev.~Lett.} {\bf #1} (19#2) #3}


\global\nulldelimiterspace = 0pt



\def\frac#1#2{{{#1} \over {#2}}\,}  
\def\hf{{1\over 2}}



\def\Dsl{\hbox{/\kern-.6700em\it D}} 
\def\dsl{\hbox{/\kern-.5300em$\partial$}}
\def\pxpsl{\hbox{/\kern-.5600em$p$}}
\def\ssl{\hbox{/\kern-.5300em$s$}}
\def\epssl{\hbox{/\kern-.5100em$\epsilon$}}
\def\delsl{\hbox{/\kern-.6300em$\nabla$}}
\def\lxpsl{\hbox{/\kern-.4300em$l$}}
\def\elxpsl{\hbox{/\kern-.4500em$\ell$}}
\def\kxpsl{\hbox{/\kern-.5100em$k$}}
\def\qxpsl{\hbox{/\kern-.5000em$q$}}
\def\sla#1{\raise.15ex\hbox{$/$}\kern-.57em #1}



\def\roughly#1{\mathrel{\raise.3ex\hbox{$#1$\kern-.75em\lower1ex\hbox{$\sim$}}}}





\def\Scl{{\cal L}}

\def\Scy{{\cal Y}}


\def\ssl{{\sss L}}







\overfullrule=0pt

\def\a{\alpha}

\def\rht{{\sss R}}
\def\lft{{\sss L}}
\def\gwk{$SU(2)_{{\sss L}} \times U(1)_{{\sss Y}}$}

\def\leff{\Scl_{\rm eff}}

\def\sw{s}
\def\cw{c}

\def\za{{\sss Z}\gamma}

\def\rht{{\sss R}}
\def\lft{{\sss L}}

\def\ww{{\sss W}}
\def\aa{{\gamma}}
\def\zz{{\sss Z}}
\def\za{{\sss Z\gamma}}

\def\gl{g_\lft}

\def\gwk{$SU_\lft(2) \times U_{\sss Y}(1)$}

\def\Mz{M_{\sss Z}}
\def\Mw{M_{\sss W}}

\def\lten{L_{{\rm 10}}}


\rightline{UTTG-04-95}
\rightline{NUHEP-TH-95-3}
\rightline{hep-ph/9504272}
\vskip .1in

\title
\centerline{Can light Goldstone boson loops}
\centerline{counter the}
\centerline{`S-argument' against Technicolor?}
\endtitle

\authors
\centerline{Sean Fleming}
\vskip .1in
\centerline{\it Department of Physics and Astronomy}
\centerline{\it Northwestern University, Evanston, IL 60208}
\vskip .15in
\centerline{Ivan Maksymyk}
\vskip .1in
\centerline{\it Theory Group, Department of Physics}
\centerline{\it University of Texas, Austin, TX 78712}
\endauthors

\abstract
We examine the oblique correction phenomenology of
one-family Technicolor with light pseudo-Goldstone
bosons.   From loop calculations based on a gauged chiral
lagrangian for Technicolor, we are lead to
conclude that
even though loops with light Goldstone bosons give
a negative contribution to $S$ measured at the $Z$-pole, this effect is not
sufficiently large to unambiguously
counter the `S-argument' against one-family Technicolor.
This result cannot be guessed \apriori, but must
be explicitly calculated.
Our analysis entails
an extended version of
the $STU$ oblique parametrization of Peskin and Takeuchi.
In principle, this
extended formalism ($STUVWX$) must be used when
there are light new particles in loops.
\medskip \leftline{PACS number(s): 12.15.Lk 12.60.Nk}
\endabstract


\vfill\eject
\section{Introduction}

\ref\pt{M.E. Peskin and T. Takeuchi, \prl{65}{90}{964};
\prd{46}{92}{381}.}

\ref\stuvwx{I. Maksymyk, C.P. Burgess, D. London, \prd{50}{94}{529};
S. Fleming, I. Maksymyk, report \# UTTG-10-95, NUHEP-TH-95-04.}
\ref\stuvwxfit{C.P. Burgess, S. Godfrey, H. K\"onig, D. London, I. Maksymyk,
\plb{326}{94}{276}.}

The precision $e^+e^-$ collision data currently being
collected will allow for a real probing of electroweak
radiative corrections and physics beyond the
Standard Model.
One method for parametrizing such effects
is the $STU$ oblique
formalism of Peskin and Takeuchi \pt, which can be used to
encode the effects of new physics electroweak gauge-boson self-energies
when these self-energies can be effectively
expressed as linear functions of $q^2$,
\eq\label\linear\Pi(q^2) = \Lambda^2
\left(  a_0 + a_1 \; {q^2\over\Lambda^2} +
O\left({q^4\over\Lambda^4}\right) \right)\eeq
with $a_0$ and $a_1$ some constants.
The $STU$ approximation
is valid when the new physics scale $\Lambda$
is much greater than the scale at which experiments are performed,
\ie\ the $Z$-pole and below.
The set of three parameters $S$, $T$ and $U$ has
recently been extended, in \stuvwx\ \stuvwxfit\ ,
to the case of light new physics,
for which the self-energies would \apriori\ be some general, complicated
functions of $q^2$.
The extended version of the
formalism involves the six parameters
$S$, $T$, $U$, $V$, $W$ and $X$.  In principle, the extended version
must be used if loop contributions to
the oblique corrections entail light new particles with masses
in the range $\sim\!\Mz$ or less.
The essence of the $STUVWX$ formalism is that
the total theoretical expression for any of the
precision electroweak observables measured
at $q^2 \approx 0$, $q^2 = \Mz^2$ or $q^2 = \Mw^2$
can be expressed as
a standard model prediction plus some linear combination
of S through X.  It turns out, moreover, that all $Z$-pole
observables can be expressed in terms of only
two parameters, $S'$ and $T'$, which are linear combinations
of $S$ through $X$.

\ref\technicolor{See the review article
E. Farhi, L. Susskind, {\it Phys. Report} {\bf 74}
(1981) 277 and references therein.}
\ref\phenoreviews{See the following review articles for a thorough discussion
of the phenomenology of Technicolor: K. Lane, hepph/9401324; hepph/9409304,
report \# BUHEP-94-24;
S.F. King, hepph/9406401, to appear in {\it Reports on
Progress in Physics.}}
\ref\at{ For a discussion of notion that
$S$ has typically been considered to be positive in Technicolor, see
T. Appelquist and J. Terning, \plb{315}{93}{139} and references therein.}

It is of interest to apply oblique correction formalisms
to models of dynamical symmetry breaking such as
Technicolor \technicolor , as this type of new physics couples
most strongly to gauge bosons and therefore essentially
generates oblique effects.
It is well known that, only a few years ago,
oblique correction considerations hinging on the parameter $S$
tended to rule out certain models of
Technicolor \phenoreviews\ \at . Least mean square fits involving the three
parameters $S$, $T$, and $U$ suggest that the measured value of $S$ is
consistent with zero, or even slightly negative, while theoretical calculations
determined $S$ to be large and positive. For example the
logarithmically divergent part of the one loop chiral lagrangian contibution
to $S$ is typically positive in Technicolor theories. In addition to this ``low
-energy'' piece there is a ``high-energy'' contribution which, when calculated
by scaling the parameters of the QCD chiral lagrangian, is also positive.

The $S$-argument against Technicolor was countered in
\at , where it was pointed out that the high-energy contribution determined
from scaling the parameters of the QCD chiral lagrangian represents an upper
bound, and that other methods used to estimate this contribution result in a
smaller or negative value for the high-energy piece. The authors of \at\
na\"{\i}vely estimate the high-energy contribution by calculating the one
loop
technifermion diagrams, and find that, after adding it
to the low-energy piece, the $S$-argument against Technicolor can be
invalidated. Thus, ref.~\at, entitled
``Revenge of the one-family Technicolor
models," re-established the possible phenomenological
viability of this model.

\ref\bb{P. Bamert and C.P. Burgess, hepph-9407203, report \# McGill-94/27,
NEIP-94-005, to appear in {\it Zeit. f\"ur Phys.}}

The calculations in the present article were embarked upon in
hope of further legitimizing Technicolor.
Our point of departure was the idea that, strictly speaking,
the results of a fit of the
three parameter set $STU$ to experimental data
can only be applied
when the physical Goldstone bosons in Technicolor are thought to be heavy.
We therefore set out to explore the possibility that some of them
are light (but just heavy enough to have
so far escaped direct detection),
and to determine whether,
in such a scenario, the theoretical values of the
new parameters $V$, $W$ and $X$
can be as large as various estimates of $S$.  If so,
the parameter $S'= S + 4(\cw^2 - \sw^2) X + 4 \cw^2 \sw^2 V$ observed at the
$Z$-pole might be consistent
with experiment.
Then one might say that
the $VWX$-argument undoes the original $S$-argument against Technicolor.

This article is organized as follows.  In Section 2, we
review how the $STU$ formalism can be extended to the
case of light new physics.  The extended formalism entails
the six parameters $STUVWX$.  In Section 3, we review the gauged chiral
lagrangian (which is
an effective lagrangian for Technicolor) and calculate
the one-loop oblique corrections, paying close attention to the sign of
$S'$, and to the ramifications of loops involving light Goldstone bosons.
We conclude in Section 4.

\section{$STUVWX$ Formalism}

\subsection{Extending the $STU$ Parameter Set}

The $STU$ formalism of Peskin and Takeuchi \pt\ provides an elegant means of
parametrizing new physics effects on electroweak observables, when
the new physics couples most strongly to gauge bosons (\ie\ oblique
corrections). This formalism allows us to write a wide range of observables
as a standard model prediction plus some linear combination of the three
parameters $S$, $T$ and $U$. The $STU$ parametrization is based
explicitly on the
assumption that new physics is heavy, and that
new physics contributions to gauge-boson self-energies are therefore
linear functions of $q^2$, \ie\ of the form of eq.~\linear.

If the heavy new physics assumption is dropped, the gauge-boson
self-energies have some complicated dependence on $q^2$ that cannot
be adequately expressed using the first few terms of a Taylor expansion.
Nonetheless, since precision observables are associated
only with the scales
$q^2 \approx 0$, $q^2 = \Mz^2$ or $q^2 = \Mw^2$, it turns out that
it is possible
in practice to parametrize oblique effects due to light new physics
in terms of only six parameters $S$,
$T$, $U$, $V$, $W$ and $X$. These are defined as
\stuvwx\ \stuvwxfit\
\eqa
\label\ssdef\alpha S =&
 - 4 \sw^2 \cw^2\widehat\Pi_\aa(0)  +
 {4 \sw^2 \cw^2\over\Mz^2}\left(\Pi_\zz(\Mz^2) - \Pi_\zz(0)\right)\eolnn
& \qquad - 4 (\cw^2 - \sw^2 )\sw\cw \; \widehat\Pi_\za(0) \eol
   \alpha T =& \;{\Pi_\ww(0)\over M^2_W} - {\Pi_\zz(0)\over\Mz^2} \eol
   \alpha U =&  - 4 \sw^4\widehat\Pi_\aa(0) +
    {4 \sw^2\over M^2_W}\left(\Pi_\ww(M^2_W) - \Pi_\ww(0)\right)\eolnn
  & \qquad - {4 \sw^2 \cw^2\over\Mz^2}\left(\Pi_\zz(\Mz^2)
   - \Pi_\zz(0)\right)  -
    8\cw \sw^3 \widehat\Pi_\za(0)  \eol
\label\vvdef\alpha V =& \Pi_\zz'(\Mz^2) - \left[ \, {\Pi_\zz(\Mz^2) -
\Pi_\zz(0)
\over \Mz^2}  \, \right] \eol
\alpha W =& \Pi_\ww'(M^2_W) - \left[ \, {\Pi_\ww(M^2_W) - \Pi_\ww(0) \over
M^2_W}  \, \right] \eol
\label\xxdef\alpha X =& - \sw \cw \left[ \, \widehat\Pi_\za(\Mz^2) -
\widehat\Pi_\za(0)
 \, \right] \eeol
\eeq
where $\widehat\Pi(q^2)\equiv \Pi(q^2)/q^2$, and where $\Pi'(q^2)$ denotes
the ordinary derivative with respect to $q^2$.
The $V$, $W$ and $X$
are intentionally defined so that they vanish
when the self-energies are linear functions of $q^2$ only,
in which case the $STU$ parametrization is exactly recovered.

We now illustrate how the above parameters appear
in expressions for observables.
First consider the low-energy neutral current
asymmetries, which depend
only on an effective $\sin^2{\Theta_{{\sss W}}}$
evaluated at $q^2\approx 0$. Just as in the Peskin-Takeuchi
parametrization, this quantity
is given by
\eq\label\sinlow
\sw^2(0)_{{\rm eff}} = \sw^2(0)^{{\sss SM}}_{{\rm eff}}
+ {\alpha S \over 4 ( \cw^2 -  \sw^2)} -
{ \cw^2\sw^2 \; \alpha T \over \cw^2 - \sw^2} \eeq
where
$\sw^2(0)^{{\sss SM}}_{{\rm eff}}$ is the standard model
prediction for some given asymmetry, and where the particular linear
combination of $S$ and $T$ is common to all
asymmetries measured at $q^2\approx 0$.

However, as to the
$Z$-pole neutral current
asymmetries such as $A_{{\sss LR}}$ and
$A_{{\sss FB}}$, the oblique corrections to
the effective $\sin^2{\Theta_{{\sss W}}}$ at the Z-pole are
given by
\eq\label\sinhigh
\sw^2(\Mz^2)_{{\rm eff}} = \sw^2(\Mz^2)^{{\sss SM}}_{{\rm eff}}
+ {\alpha S \over 4 ( \cw^2 -  \sw^2)} -
{ \cw^2\sw^2 \; \alpha T \over \cw^2 - \sw^2} + \alpha X.\eeq
Here, the parameter $X$ represents a supplementary $Z$-pole
effect, defined in eq.~\xxdef.

In the full $STUVWX$ formalism, the neutral current vertex at the $Z$-pole
is multiplied by an overall oblique correction factor
$(1 + \alpha T + \alpha V)$.  Thus,
for example, the width of Z-decay
to neutrinos is given by
\eq\label\Zdecay
\Gamma(Z\rightarrow \nu\overline{\nu}) =
\Gamma(Z\rightarrow \nu\overline{\nu})^{{\sss SM}}
( 1 + \alpha T + \alpha V ).\eeq
In studying eqs.~\sinlow , \sinhigh\ and \Zdecay , one sees
that when one drops the assumption of heavy new physics and, with it,
the corresponding linear approximation,
it is a simple matter to systematically incorporate
the new physics
oblique effects into expressions for $Z$-pole observables.

Similarly, the width of W-decay to a single lepton-neutrino pair
is given by
\eq\label\Wdecay
\Gamma(W\rightarrow l\overline{\nu}) =
\Gamma(W\rightarrow l\overline{\nu})^{{\sss SM}}
  \left( 1 - { \alpha S \over 2 ( \cw^2 -
  \sw^2)} + { \cw^2\; \alpha T \over (\cw^2 - \sw^2)}  +
{\alpha U \over 4 \sw^2} +
\alpha W \right) .
\eeq

Finally, note that in the $STUVWX$ formalism,
the mass of the $W$-boson is given
by a formula identical to that arising in the
$STU$ formalism, namely
\eq\label\wmass
\Mw^2 =
(\Mw^2)^{{\sss SM}}
\left(1 - {\alpha S\over 2(\cw^2 - \sw^2)} +
            {\alpha \cw^2 T\over \cw^2 - \sw^2} +
            {\alpha U \over 4 \sw^2} \right) .
\eeq

\ref\lll{L. Lavoura, L.F. Li, \prd{48}{93}{234}.}

\ref\pl{P. Langacker, Talk given at 22nd INS Symposium on
Physics with High Energy Colliders, Tokyo, March, 1994, hepph-9408310.}
\ref\lepfit{P. Bamert, C.P. Burgess, I. Maksymyk, report \# UTTG-09-95.}

\subsection{Oblique Parameters for Z-pole Measurements}

The formalism described above is the most natural extension of the $STU$
parameterization, though it does have the disadvantage that $X$ and $V$
appear in the expressions for $Z$-pole observables.
It is, however, possible to cast the
oblique corrections to all $Z$-pole observables in terms of
only two parameters, which, following \bb , we may conveniently define as
\eqa\label\bamertandburgess
S' = & \; S + 4 (\cw^2 - \sw^2 ) X + 4 \cw^2 \sw^2 V \eolnn
T' = & \; T + V. \eeol\eeq
\ref\ab{G. Altarelli, R. Barbieri, \plb{253}{91}{161};
G. Altarelli, R. Barbieri, S. Jadach,
\npb{369}{92}{3}.}
The effective vertex for neutral currents at the $Z$-pole is now
given by
\eq\label\effzpole
i\Lambda^\mu_{{\rm nc}}(q^2\!=\!\Mz^2)
= - i \; {e\over \sw\cw} \; ( 1 + {1 \over 2}\alpha T') \;
\gamma^\mu
\left[ I_3^f \gamma_\lft
- Q^f \left( \sw^2 +   {\alpha S' \over 4 ( \cw^2 -  \sw^2)} -
{ \cw^2\sw^2 \; \alpha T' \over \cw^2 - \sw^2}  \right) \right] .
\eeq
So, in confronting
some model of light new physics with $Z$-pole data, one
would calculate $S'$ and $T'$ rather than $S$ and $T$.
The $\epsilon$ parameters of Altarelli and Barbieri \ab\ are
connected to these parameters by
$\epsilon_1 = \alpha T'$ and $\epsilon_3 = \alpha S'/(4\sw^2)$.
With $S'$ and $T'$ defined this way, the low-energy neutral-current observables
now depend on $S', T', V$, and $X$; the $W$-mass depends on
$S', T', U, V$, and $X$.

The results of fits to precision data for the parameters $STU$
can be found in \pl , and for $STUVWX$ in \stuvwxfit.
Fits to the most recent LEP and SLC data (Winter 1995) are presented in
\lepfit, the result being
\eqa\label\fitresults
S' = & \; - 0.20 \pm 0.20 \eolnn
T' = & \; - 0.13 \pm 0.22 \eol
\alpha_{{\rm s}}(\Mz) = & \; 0.127 \pm 0.005 \eeolnn\eeq

\section{Calculation of $S$ through $X$ in One-Family Technicolor}

\subsection{Gauged Chiral Lagrangian for Technicolor}

\ref\peskin{Micheal E. Peskin \npb{175}{80}{197}.}
\ref\nonlinear{
A.C. Longhitano, \prd{22}{80}{1166};
\npb{188}{81}{118};
S. Chadha, M.E. Peskin, \npb{185}{81}{61};
F. Feruglio, \ijmp{8}{93}{4937}.}
\ref\pr{M.E. Peskin, R. Renkin, \npb{211}{83}{93}.}
\ref\ss{M. Soldate, R. Sundrum \npb{340}{90}{1}.}
\ref\rg{M. Golden, L. Randall \npb{361}{91}{3}.}

Our approach
consists of using an effective lagrangian (the
gauged chiral lagrangian
\peskin\ \nonlinear\ \pr\ \ss\ \rg\ )
to calculate one-loop contributions to the
self-energies of the electroweak gauge bosons.

Let us consider
the ``one-family" model of Technicolor.  In this model,
a chiral symmetry $SU(8)_\lft\times SU(8)_\rht$
is realized on a set of technifermions of eight flavours:
$(U^{\a}_{r},D^{\a}_{r}$, $U^{\a}_{b},D^{\a}_{b}$,
$U^{\a}_{g},D^{\a}_{g}$, $E^{\a},N^{\a})$.
There is a flavour of technifermion
for each distinguishable member of a one-family representation
of the usual gauge group
$G\equiv SU(3)_{{\sss C}}\times SU(2)_\lft\times U(1)_{{\sss Y}}$.
`$\a$' indexes Technicolor, a new color-like force.
The technifermions have the same
quantum numbers under $G$ as the corresponding ordinary fermions.
It is assumed
that ordinary fermions are singlets under Technicolor.
The new color-like Technicolor force becomes strong
at some scale $\Lambda_{{\sss TC}}$ in the TeV range, resulting in the
breaking of the chiral symmetry to
$SU(8)_{{\sss V}}$ and in the
formation
of Goldstone bosons, called ``technipions," which
are bound states of two technifermions.
This is exactly analogous to
the formation of pions and the breaking of chiral symmetry
in ordinary hadronic physics.

Following
\nonlinear\ \ss\ and \rg,
we define
\eq\label\u
U = \exp{{ i 2 X_i \Pi_i \over v }}
\eeq
where $\Pi_i$ are the 63 technipion fields associated with
the breaking of the chiral symmetry and where $X_i$ are
the 63 \ $8\!\times \!8$ traceless hermitian matrices that generate $SU(8)$,
normalized so that
\eq\label\normalization
Tr \left[ X_i X_j \right] = \; \hf\; \delta_{ij}.
\eeq
The gauged chiral lagrangian is written as
\eq\label\gaugedchirallag
\Scl = \Scl_{{\rm kin}} + \Scl'
\eeq
where the most important terms are found in
\eq\label\kinetic
\Scl_{{\rm kin}} = {v^2 \over 4}Tr\left[ (D^\mu U)^\dagger \; D_\mu U \right],
\eeq
and where $\Scl'$, contains a set of \gwk-invariant terms, including
terms up to some
given order in derivatives.
The \gwk\ covariant derivative is
given by
\eq\label\covariantderivative
D^\mu U = \partial^\mu U - i g \sqrt{N_d} W^\mu_i T_i \, U
+ ig'\sqrt{N_d} (U T_3 B^\mu - \left[ \Scy, U \right] B^\mu)
\eeq
where $N_d$ is the number of technidoublets,  where
\eq\label\ithree
T_i = \; {1 \over \sqrt{N_d}} \; \tau_i\otimes I_{N_d} ,
\eeq
and where, for one-family Technicolor (with $N_d = 4$), we have
\eq\label\why
\Scy =    {1\over 2}
\pmatrix{{1\over3}\tau_0 &              &                &  \cr
                         &{1\over3}\tau_0&               &  \cr
                         &             &{1\over3}\tau_0 &   \cr
                         &             &                 & - \tau_0 \cr } .
\eeq
We define
$\tau_0\equiv {1\over 2} I_2$ and $\tau_i\equiv {1\over 2} \sigma_i$.
The explicit $\sqrt{N_d}$ displayed in eq.~\covariantderivative\
assures that, for example, the mass of the $W$-boson works out
to $\Mw^2 = {1\over 4} g^2 N_d v^2$ as required.
The gauged chiral lagrangian is invariant under the local
\gwk\ transformation in which the Goldstone bosons transform according to
\eq\label\localewk
U\rightarrow \;
e^{i(\beta\Scy + \alpha_iT_i)} \;
U \;e^{-i\beta(\Scy + T_3)} ,
\eeq
and in which the gauge bosons transform according to
the usual Yang-Mills transformation rule.
Of the 63 Goldstone bosons, three are eaten, leaving 60
physical pseudo-scalars in the theory.

\ref\ht{B. Holdom, J. Terning, \plb{247}{90}{88}.}
\ref\gl{J. Gasser, H. Leutwyler, \npb{250}{85}{465};
G. Ecker, J. Gasser, H. Leutwyler, A. Pich, E. de Rafael, \plb{223}{89}{425};
A. Bay et al., \plb{174}{86}{445};
S. Egli et al., \plb{175}{86}{97}.}

\subsection{Calculation of ``High-Energy''
Contribution by Scaling of QCD Results}

Before proceding with our loop calculations,
we will look at the sector $\Scl'$ appearing in eq.~\gaugedchirallag .
This sector consists of an expansion in derivatives of all the
locally \gwk -invariant terms that one can construct from gauge-boson
and Goldstone boson fields.
Among the interactions included in $\Scl'$ is the operator
$B_{\mu\nu}W^{\mu\nu}_{i}
Tr \left[ U^\dagger T_{3} U T_{i} \right]$, which gives
a ``high-energy'' contribution to the oblique parameter $S$.
The operator's
coefficient is defined via,
\eq\label\lteneffqcd
\leff^{{\sss QCD}} = \; \lten^{{\sss QCD}}  g g' B_{\mu\nu}W^{\mu\nu}_{i}
Tr \left[ U^\dagger \tau_{3} U \tau_{i} \right]\; + \dots,
\eeq
where the experimental value
\eq\label\value
\lten^{{\sss QCD}}(\!\Lambda_{\sss QCD}\!) = \; - 5.4 \pm 0.3 \times 10^{-3}
\eeq
is determined from measurments of the pion charge and the
decay $\pi\rightarrow e\nu\gamma$ \gl.
To find the correct normalization of this operator in our conventions
for technicolor with $N_d$ doublets,
we note that the contribution to a gauge-boson two-point function
is $ \lten^{{\sss QCD}} N_d N_{{\sss TC}}/N_{{\sss QCD}}$.
This direct physical association requires that we write
\eq\label\ltenefftc
\leff^{{\sss TC}} = \; N_d
\; { N_{{\sss TC}}\over N_{{\sss QCD}} }\;
\lten^{{\sss QCD}}  g g' B_{\mu\nu}W^{\mu\nu}_{i}
Tr \left[ U^\dagger T_{3} U T_{i} \right] \; + \dots .
\eeq
The gauge boson two-point function
embedded in the above equation is
\eq\label\twopoint
\leff^{{\sss TC}} = \; {N_d \over 2}
\; { N_{{\sss TC}}\over N_{{\sss QCD}} }\;
\lten^{{\sss QCD}}  g g' B_{\mu\nu}W^{\mu\nu}_{3} \; + \dots   .
\eeq

Since $S$ is generically associated with $-32\pi\sw\cw/e^2$ times
the coefficient of the $B_{\mu\nu}W^{\mu\nu}_{3}$ term,
we have
\eq\label\svalue
S(\Lambda_{{\sss TC}}) = \;
- 16 \pi \; {N_d N_{{\sss TC}}\over N_{{\sss QCD}} } \;
\lten^{{\sss QCD}}(\Lambda_{{\sss QCD}}) \sim +1   .
\eeq
It is this large positive ``high-energy'' contribution,
combined with the positive logarithm that
is calculated in the next subsection, that, at the outset,
renders the model unviable.  (See eq.~\fitresults.)
In ref.~\at, however, it was
pointed out that the high-energy contribution can be
estimated na\"{\i}vely by simply calculating
the technifermion loops, yielding a result that can be as low as $-0.2$.

Finally, the entire phenomenological value of the measured
quantity $S(\Mz)$ is given by
\eq\label\totalvalue
S(\Mz) = \; S(\Lambda_{{\sss TC}}) + S
\eeq
where $S$ henceforth refers to the the contribution obtained by
calculating gauge-boson self-energies involving physical
Goldstone boson loops.
Such loops are calculated in the next subsection.
The logarithmically divergent parts of
$S$ give the renormalization group scaling of
$S(\mu)$ from $\Lambda_{{\sss TC}}$ down to $\Mz$.

\subsection{Goldstone Boson Loop Calculations}

The interactions pertinent to our
one-loop calculations are
the Goldstone-Goldstone-gauge-boson ($GGg$) and
Goldstone-Goldstone-gauge-gauge-boson ($GGgg$) interactions
embedded in eq.~\kinetic.
The relevant Feynman rules are given in Fig.~1.
Such couplings contribute to the gauge boson self-energies
through the one-loop diagrams shown in Fig.~2.

\ref\evans{N. Evans, \prd{49}{94}{4785}.}

The one-loop contributions to oblique corrections
in the gauged chiral lagrangian
have been studied in \pr\ \rg\ \evans.
In refs. \pr\ and \rg, the
{\it (logarithmically) divergent} parts of
various electroweak observables were
calculated only.
Since the divergent parts of the self-energies
turn out to be linear functions of $q^2$, these analyses fit
into the framework of the $STU$ formalism.

The author of ref. \evans , on the other hand,
explicitly considered the possibility of light new particles,
and thus adopted the $STUVWX$ formalism.  In performing
one-loop calculations with a degenerate triplet of
Goldstone bosons, this author was concerned only
with the finite parts of the gauge-boson self-energies,
and as a result, did not display the divergent parts
(all of which all reside in the parameter $S$).
Moreover, in ref. \evans, it was not asked whether the $VWX$-argument
could help undo the $S$-argument against Technicolor.

\medskip

\leftline{$\bullet$ Goldstone Boson Isotriplets}

Calculating the loop contributions from a degenerate non-self-conjugate
isotriplet of Goldstone bosons (and its conjugate triplet),
we obtain the following self-energy
pieces:
\eqa\label\selfenergies
\Pi_\aa(q^2) & = e^2 (2 + 3 y^2) (I(q^2) - 2 J)\eolnn
\Pi_\za(q^2) & = e^2 {(1-2\sw^2 - 3 \sw^2 y^2)\over  \sw\cw}
(I(q^2) - 2 J)\eolnn
\Pi_\zz(q^2) & = e^2
\left(
{(1 - 2 \sw^2)^2 + 6 s^4 y^2)\over 2 \sw^2\cw^2} I(q^2)
+ {2\over\cw^2}(2\cw^2 - 3 \sw^2 y^2) J
\right)\eolnn
\Pi_\ww(q^2) & = {e^2\over 2 \sw^2} I(q^2) \eeol\eeq
where $y$ is the hypercharge of the triplet,
defined through $Q = I_3 + Y$, and
where $I(q^2)$ and $J$ correspond to the contributions
from from figures 2a and 2b respectively.  They are defined as
\eqa\label\iandj
I(q^2) &= {1\over 8 \pi^2}
\left[ \! \! \left( m_\pi^2 - {q^2\over 6} \right) \! \!
\left( \! {1\over\epsilon'} + \log{{ \mu^2 \over m_\pi^2 }} \! \right)  -
\int_0^1 \! \! dx (m_\pi^2 \! - \! q^2 ( x \!  - \! x^2 ))
\log{ \! \left( \! 1 -
{q^2\over m_\pi^2}(x \! - \! x^2) \! \! \right) } \!\! \right]
\eol
J &= {1\over 16\pi^2} m_\pi^2
\left[
{1\over\epsilon'} + \log{{ \mu^2 \over m_\pi^2 }} \right]
\eeol\eeq
where ${1/\epsilon'} \equiv 2/(n-4) - \gamma + 1 + \log{4\pi}$.
We interpret the $1/\epsilon'$ coefficient as determining
the logarithmic scaling of $S$ from $\Lambda_{{\sss TC}}$
down to $\Mz$.

Using the definitions of $S-X$ given in
eqs.~\ssdef\ through
\xxdef , we obtain for the
degenerate non-self-conjugate isotriplet and its conjugate:
\eqa\label\triplettechnistuvwx
\a S &= {e^2 \over 24 \pi^2}
\log{\Lambda^2_{{\sss TC}} \over \Mz^2} +
 {\hbox{convergent pieces}}\eolnn
\a T &= 0 \eolnn
\a U {4 \pi^2 \over e^2} &=
- { 2 \sw^2\cw^2\over 3}  +  \sw^4 y^2
 - \int_0^1 dx
\left( {m_\pi^2 \over \Mw^2} - (x - x^2) \right)
\log{ \left( 1 - {\Mw^2\over m_\pi^2}(x - x^2) \right) } \eolnn
& + ((1 - 2 \sw^2)^2 + 6 \sw^4 y^2)
\int_0^1 dx
\left( {m_\pi^2 \over \Mz^2} - (x - x^2) \right)
\log{ \left( 1 - {\Mz^2\over m_\pi^2}(x - x^2) \right) }
\eolnn
\a V &= {e^2 \over 16 \pi^2 \sw^2 \cw^2}\;
((1-2\sw^2)^2 + 6 \sw^4 y^2 )
\left[
{m_\pi^2 \over \Mz^2} \int_0^1 dx
\log{ \left( 1 - {\Mz^2\over m_\pi^2}(x - x^2) \right) } + {1\over 6}
\right]
\eolnn
\a W &= {e^2 \over 16 \pi^2 \sw^2}
\left[
{m_\pi^2 \over \Mw^2} \int_0^1 dx
\log{ \left( 1 - {\Mw^2\over m_\pi^2}(x - x^2) \right) } + {1\over 6}
\right]
\eolnn
\a X &=
{e^2 \over 8 \pi^2} \; ( 1 - 2 \sw^2 - 3 \sw^2 y^2 )
\left[ \int_0^1 dx
\left( {m_\pi^2\over \Mz^2 }  - (x - x^2) \right)
\log{ \left( 1 - {\Mz^2\over m_\pi^2}(x - x^2) \right) } + {1\over 6}
\right]  .
\eolnn
{}
\eeol
\eeq
(The results for a degenerate
self-conjugate isotriplet can be obtained from the
above expressions by setting $y\!=\!0$ and dividing by two.)

Note that the logarithmic divergence in $S$ is positive, and,
as it turns out, strictly independent of the hypercharge
$y$.  Thus, no exotic values of hypercharge can be
evoked to render $S$ negative.
$T$ is exactly zero
(because of the degeneracy of the triplet).
$U$, $V$, $W$ and $X$ are finite, and therefore can be evaluated
unambiguously.  Below, we display the results for $V$ and $X$
in two interesting limits: $m_\pi = \hf \Mz$
(for which an exact expression is easily obtained)
and $m_\pi \gg \Mz$ (for which we can expand in $\Mz^2/m_\pi^2$).
The results are given, respectively, by
\eqa\label\interesting
\a V &= {e^2 \over 16 \pi^2 \sw^2 \cw^2}\;
((1-2\sw^2)^2 + 6 \sw^4 y^2)
\left( \;\;\;  - \;\;\;{1\over 3} \;\;\; ,
\;\;\; - \;\;\; {1 \over 60}\; {\Mz^2  \over m_\pi^2} \;\;\;
+ \;\;\; O({\Mz^4  \over m_\pi^4} ) \;\;\;
\right)
\eolnn
\a X &=
{e^2 \over 8 \pi^2} \; ( 1 - 2 \sw^2 - 3 \sw^2 y^2 )
\left( \;\;\;   + \;\;\;{1\over 9} \;\;\; ,
\;\;\; + \;\;\; {1 \over 60}\; {\Mz^2 \over m_\pi^2 } \;\;\;
+ \;\;\; O({\Mz^4  \over m_\pi^4} ) \;\;\;
\right) .
\eeol
\eeq
The above results for the large $m_\pi$ limit
have the peculiar feature that
the coefficient of the first term in
the Taylor expansion is surprisingly small.
Thus we see that as $m_\pi$ increases from
$\Mz/2$ to, say, $2 \Mz$, the size of $V$ or $X$
is diminished by at least one full order of magnitude!

\medskip

\leftline{$\bullet$ Goldstone Boson Isosinglets}

The
contributions to the self-energies due
to a non-self-conjugate singlet are given by
\eqa\label\singletselfenergies
\Pi_\aa(q^2) & = e^2 y^2 ( I(q^2) - 2 J )\eolnn
\Pi_\za(q^2) & = - e^2 y^2 { \sw \over \cw }
(I(q^2) - 2 J)\eolnn
\Pi_\zz(q^2) & = e^2 y^2
{ \sw^2 \over \cw^2 }
\left( I(q^2)
-  2 J \right)\eolnn
\Pi_\ww(q^2) & = 0. \eeol\eeq
With these self-energy contributions, we obtain the following
results for the parameters $S$ through $X$:
\eqa\label\singlettechnistuvwx
\a S &= - {e^2 \sw^4 y^2 \over 2 \pi^2}
\left[
\int_0^1 dx
\left( {m_\pi^2\over \Mz^2 }  - (x - x^2) \right)
\log{ \left( 1 - {\Mz^2\over m_\pi^2}(x - x^2) \right) } + {1\over 6}
\right]
\eolnn
\a T &= 0 \eolnn
\a U &= {e^2 \sw^4 y^2 \over 2 \pi^2}
\left[
\int_0^1 dx
\left( {m_\pi^2\over \Mz^2 }  - (x - x^2) \right)
\log{ \left( 1 - {\Mz^2\over m_\pi^2}(x - x^2) \right) } + {1\over 6}
\right]
\eolnn
\a V &= {e^2 \sw^2 y^2 \over 8 \cw^2 \pi^2}
\left[
{m_\pi^2 \over \Mz^2} \int_0^1 dx
\log{ \left( 1 - {\Mz^2\over m_\pi^2}(x - x^2) \right) } + {1\over 6}
\right] \eol
\a W &= 0 \eolnn
\a X &= - {e^2 \sw^2 y^2 \over 8 \pi^2}
\left[
\int_0^1 dx
\left( {m_\pi^2\over \Mz^2 }  - (x - x^2) \right)
\log{ \left( 1 - {\Mz^2\over m_\pi^2}(x - x^2) \right) } + {1\over 6}
\right] . \eeolnn\eeq
(To obtain the result for a self-conjugate singlet,
one simply sets $y$ to zero, \ie\ there is no contribution
from a self-conjugate singlet.)
The above formulae illustrate that the contributions due to a
non-self-conjugate singlet are all finite.  Interestingly,
one discovers, upon evaluation of the integral, that the
above result for $S$ is generally
negative. For $m_\pi = \Mz/2$, we have
\eq
\alpha S = \; - \; {e^2 \sw^4 y^2\over 2 \pi^2}\;\left( {1 \over 9} \right) ,
\eeq
and for $m_\pi \gg \Mz$, we have
\eq
\alpha S = \; - \; {e^2 \sw^4 y^2\over 2 \pi^2}\;
\left( {1 \over 60}\; { \Mz^2\over m_\pi^2} \right) .
\eeq
This negative value could be taken as a reassuring sign
if one wanted to further
establish the phenomenological feasibility of Technicolor.
However, it must be appreciated that
of the 60 physical Goldstone
bosons in one-family Technicolor, only three pairs of particles
(the coloured isosinglets designated as $T_c$
and $\overline{T}_c$ in \technicolor) are
non-self-conjugate singlets. The great majority of the
Goldstone bosons are
arranged in triplets, and therefore the negative $S$ contributions
from the few non-self-conjugate singlets cannot
effectively counter the positive contributions from the
many triplets.

\subsection{Numerical Estimates}

Estimates for the masses of the various Goldstone bosons
are presented in \technicolor.   Most
of these particles (those designated as $T_c^i$,
$\overline{T}_c^i$ and $\theta_a^i$, constituting a
total of 14 triplets) are expected to have masses of roughly
$m_\pi = 200$ GeV.
Taking this value for $m_\pi$ and
taking $\Lambda_{{\sss TC}} \approx 1$ Tev, one obtains
for an individual (self-conjugate) triplet
\eqa\label\stuvwxvalues
& S =  \;\; {1 \over 12 \pi}
\log{{\Lambda_{{\sss TC}}^2\over \Mz^2}} + {\hbox{convergent pieces}}
\;\;\; \sim \;\;\;  O(0.1)\eolnn
& U, V, W, X \; \sim \; O(0.0001) .   \eeol\eeq
The essential result is therefore that, for a triplet of mass
$m_\pi = 200$ GeV, the chiral loop
contribution to $S$ is significantly larger than the
contribution to the other parameters.

Let us next examine the case of lighter Goldstone bosons.
In one-family Technicolor, there does exist one (self-conjugate)
triplet of particularly light
physical Goldstone bosons, the $P_i$,
with mass estimated to be less than 100 GeV \technicolor .
To estimate the most dramatic possible contribution of this triplet, let us
assume (as in eq.~\interesting\ ) that
$m_\pi \!=\!M_Z/2$, \ie\ that the technipions are as light
as possible while
being just out of reach of direct detection.  In this case,
evaluation of eqs.~\triplettechnistuvwx\ and \interesting\ gives
\eqa\label\stuvwxvaluesp
S = & \;\; {1\over 12 \pi}
\log{{\Lambda_{{\sss TC}^2}\over \Mz^2}} + {\hbox{convergent parts}}
\;\;\; \sim \;\;\; O(0.1) \eolnn
T = & \; 0 \eolnn
U = & -0.006 \eolnn
V = & - 0.02  \eol
W = & -0.02   \eolnn
X = & +0.005 .  \eeolnn\eeq
{}From eq.~\stuvwxvaluesp ,
it can be appreciated that the oblique quantity
which is measured at the $Z$-pole,
$ S'\equiv S + 4 (\cw^2 - \sw^2 ) X + 4 \cw^2 \sw^2 V $,
does not receive an appreciable
negative contribution from the $V$ and $X$ terms.
Therefore, it appears that the $VWX$-argument does not help to
undo the $S$-argument against one-family Technicolor.
This result cannot be guessed \apriori, but must be determined
through explicit calculation.
Surprises and ``conspiracies"
can occur in these
calculations. For example, it has been noticed
\lll\ that in extensions of the Standard Model involving
doublets of fermions or multiplets of scalar bosons, the
photon-$Z$ self-energy is proportional to the very
small quantity ${1\over 4} - s^2$, so that
$X$ is by chance much smaller than the other parameters; in
the present calculation, however, this particular combination
did not arise naturally.  Moreover, division by
$s^2$ can give rise to an important enhancement, and such
an enhancement might well have affected our results
qualitatively.

It is interesting to note that there exists a small negative
contribution to $T'$, due to the additional $V$ piece
in $T' = T + V = -0.02$.
Thus, we find that even a perfectly degenerate
triplet of scalars yields a non-zero (and negative!)
contribution to the effective $\rho$ parameter measured at
the $Z$-pole.    This result is not without phenomenological
pertinence: for example, a 10 GeV deviation of the top
mass from a fiducial value of 178 GeV gives a change
in $T$ of the same order, namely $\approx \pm 0.06$.

\section{Conclusion}

The `S-argument' against
Technicolor hinges on the fact that the value of $S$ calculated in a one-family
Technicolor model is large and positive, while the experimental measurements
of $S$ at the $Z$-pole are consistent with zero. In a one-family Technicolor
model with light pseudo-Goldstone bosons the parameter that is measured at the
$Z$-pole is $S'$, where $ S'\equiv S + 4 (\cw^2 - \sw^2 ) X + 4 \cw^2 \sw^2 V
$.
Thus it is clear that if either $X$ or $V$ are large and negative the
calculated
value of $S'$ can be consistent with the experimental data. The result of such
a
calculation can not be guessed \apriori.
We have calculated the parameters $STUVWX$ in a one-family
Technicolor model with light pseudo-Goldstone bosons, and found that the values
of $V$, and $X$ do not contribute significantly to $S'$. Hence one-family
Technicolor models with light psuedo-Goldstone bosons can not counter the
`S-argument' against Technicolor.

Though the values
of $V$, and $X$
do not play a predominant role. One ought to keep in mind though
that, as is
discussed in refs.~\stuvwx, \bb\ and \lll,
there do indeed exist models of
new physics in which the extended set of parameters may well be relevant.
Thus, it is possible that the $STUVWX$ parameter set might
one day participate in untangling some
signal of physics beyond the Standard Model.

\bigskip

\noindent {\bf Acknowledgements:} The work of I.M. was supported by Robert A.
Welch Foundation, NSF Grant PHY90-09850, and by NSERC of Canada. The work of
S.F. was supported in part by the U.S. Department of Energy, Division of High
Energy Physics under Grant DE-FG02-91-ER40684.
The authors would like to thank Nick Evans, Probir Roy
and John Terning for
helpful comments.

\listrefs

\bye